\documentclass[aps, prb, twocolumn, floatfix, superscriptaddress]{revtex4-1}  
\usepackage[pdftex]{graphics}
\usepackage[usenames]{color}
\usepackage{microtype}
\usepackage{amsmath}
\usepackage{amssymb}
\usepackage{xspace}
\usepackage{footnote}
\usepackage{relsize}

\newcommand{\ionox}{_{\text{\tiny{IO}}}}
\newcommand{\ionhyd}{_{\text{\tiny{IH}}}}

\newcommand{\lith}{Li$^{+}$\xspace}
\newcommand{\fluor}{F$^{-}$\xspace}

\newcommand{\tst}{^{\ddag}}
\newcommand{\cent}{_{n}}
\newcommand{\deltag}{$\Delta G_{\text{hyd}}$}
\newcommand{\dgstan}{$\Delta G^{\,\ensuremath{\mathlarger{\mathlarger{-\kern-8pt{\circ}\kern-8pt-}}}}_{\text{hyd}}$}
\newcommand{\standstate}{^{\,\ensuremath{\mathlarger{\mathlarger{-\kern-8pt{\circ}\kern-8pt-}}}}}

\makeatletter
\if@twocolumn
 
\else
 
\fi
\makeatother

\makeatletter
\newcommand\footnoteref[1]{\protected@xdef\@thefnmark{\ref{#1}}\@footnotemark}
\makeatother

\begin{document}

\title{Nuclear quantum effects in water exchange around lithium and fluoride ions}

\author{David M. Wilkins}
\affiliation{Physical and Theoretical Chemistry Laboratory, 
University of Oxford, South Parks Road, Oxford OX1 3QZ, UK}

\author{David E. Manolopoulos}
\affiliation{Physical and Theoretical Chemistry Laboratory, 
University of Oxford, South Parks Road, Oxford OX1 3QZ, UK}

\author{Liem X. Dang}
\affiliation{Physical Sciences Division, Pacific Northwest National Laboratory, Richland, Washington 93352, USA}
\begin{abstract}
We employ classical and ring polymer molecular dynamics simulations to study the effect of nuclear quantum fluctuations on the structure and the water exchange dynamics of aqueous solutions of lithium and fluoride ions. While we obtain reasonably good agreement with experimental data for solutions of lithium by augmenting the Coulombic interactions between the ion and the water molecules with a standard Lennard-Jones ion-oxygen potential, the same is not true for solutions of fluoride, for which we find that a potential with a softer repulsive wall gives much better agreement. A small degree of destabilization of the first hydration shell is found in quantum simulations of both ions when compared with classical simulations, with the shell becoming less sharply defined and the mean residence time of the water molecules in the shell decreasing. In line with these modest differences, we find that the mechanisms of the exchange processes are unaffected by quantization, so a classical description of these reactions gives qualitatively correct and quantitatively reasonable results. We also find that the quantum effects in solutions of lithium are larger than in solutions of fluoride. This is partly due to the stronger interaction of lithium with water molecules, partly due to the lighter mass of lithium, and partly due to competing quantum effects in the hydration of fluoride, which are absent in the hydration of lithium.
\end{abstract}

\maketitle

\section{Introduction}\label{sec:introduction}

The exchange of water around solvated ions is a key step in biochemical processes such as reactions at protein surfaces,\cite{Modig2004,Russo2005} and in geochemical processes such as the deposition of clays and minerals from aqueous solutions\cite{Casey2006,Panasci2012} and the absorption of contaminants by minerals in soil.\cite{Udeigwe2011} In all three of these examples, understanding the seemingly simple reaction

\begin{equation}
\text{I}^{\pm z}\left( \text{H}_{2}\text{O}\right)_{n} + \text{H}_{2}\text{O}^{\ast} \rightleftharpoons \text{I}^{\pm z}\left( \text{H}_{2}\text{O}\right)_{n-1}\left( \text{H}_{2}\text{O}^{\ast}\right) + \text{H}_{2}\text{O},
\end{equation}

\noindent 
is a prerequisite for understanding much more complex processes. A great deal of work has therefore been done on this water exchange reaction. In particular, the mean residence time (MRT), $\tau$, of water in the hydration shells of ions -- or equivalently its reciprocal, the exchange rate $k$ -- has been the subject of numerous experimental and computational studies.\cite{Impey1983,Merbach1987,Ohtaki1993,Rey1996,Helm2002,Helm2005,Rustad2006,Kerisit2009,Dang2013} When the exchange is slow enough, the MRT can be inferred from nuclear magnetic resonance (NMR) measurements.\cite{Luz1965,Helm2008,Balogh2008} When the exchange is faster, NMR only provides an upper bound on the residence time, which can be reduced to a 100 ps upper bound with an incoherent quasi-elastic neutron scattering experiment.\cite{Salmon1987,Salmon1989,Salmon1990}  Shorter (sub 10 ps) residence times have also recently been measured more precisely for certain anions by two-dimensional infrared spectroscopy.\cite{Moilanen2009,Minbiao2010}

One popular computational route to the MRT was suggested by Impey \emph{et al.}\cite{Impey1983}: $\tau$ can be calculated as the decay time of a residence correlation function. This method, often denoted ``direct simulation'', relies on being able to observe exchange events in a simulation, so cannot be used if $\tau$ is too long. However, for most monovalent and monatomic ions, $\tau$ is on the order of 10 ps or less\cite{Impey1983,Koneshan1998,Helm2005} (with the exception of \lith, for which values as high as 400 ps have been reported\cite{Lee1996}), and the calculation of the MRT in this way is generally viable. A similar hydrogen-bond correlation function, introduced by Luzar and Chandler,\cite{Luzar1996,Luzar1996a} has also been used to calculate residence times of water in the first hydration shells of anions.\cite{Chandra2003,Habershon2014} An alternative method was introduced by Rey and Hynes,\cite{Rey1996,Spangberg2003a} who treated the escape of water from the first hydration shell of the ion as an activated chemical process with rate constant $k = \tau^{-1}$, calculated using the method of reactive flux. This has the advantage that there are no restrictions on the magnitude of $\tau$, allowing the study of exchanges that are too slow for the direct simulation method.\cite{Kerisit2009}

There are several other methods that have been used to calculate $\tau$: a few \emph{ab initio} studies of water exchange have found the MRT by counting the number of exchange events per unit time.\cite{Hofer2004,Messner2011} Unlike the direct simulation method, this does not require the fitting of a correlation function to an exponential decay, so may be advantageous when this fitting is problematic. Laage and Hynes\cite{Laage2008} also proposed calculating $\tau$ using the stable-states picture of chemical reactions, which is similar to the reactive flux method, but can also be used if the MRT does not reach a plateau value. Finally, Kerisit and Rosso\cite{Kerisit2009} used transition path sampling (TPS) to calculate $\tau$ for water around Na$^{+}$ and Fe$^{2+}$; this technique is more general than the reactive flux method because it does not require a pre-defined reaction coordinate.

By itself, the rate constant gives no information about the water exchange mechanism. It can, however, be used to obtain the activation volume (the difference in volume between reactant and transition state), which can be used to shed more light on the process. This is defined as

\begin{equation}\label{eq:activ_vol}
\Delta V\tst = -\beta^{-1}\left( \frac{\partial \ln k}{\partial p}\right)_{\beta},
\end{equation}

\noindent where $\beta = 1/k_{\text{B}}T$ is the inverse temperature and $p$ the pressure. $\Delta V\tst < 0$ suggests an associative mechanism (with an intermediate that has a higher coordination number than the reactant or product), and $\Delta V\tst > 0$ a dissociative mechanism (whose intermediate has a lower coordination number).\cite{Langford1966,Merbach1987} The activation volume can be probed experimentally using variable-pressure NMR,\cite{Helm2002,Balogh2008} although this technique has not been applied to solutions of monovalent ions, whose water exchange rates are too fast for water in the hydration shell to be distinguished from that in the bulk. It has been used to find activation volumes for exchange around di- and trivalent metal ions, including transition metals and lanthanides.\cite{Ducommun1980,*Hugi1985,*Hugi1987,Cossy1989}

Direct simulation can be used to calculate activation volumes for water exchange: Hermansson \emph{et al.} extracted $\Delta V\tst$ for \lith and Na$^{+}$ ions from the partial molar volumes of the species involved in the exchange reactions.\cite{Spangberg1997,Hermansson1998} Rustad and Stack computed the rate constant $k$ at multiple pressures using the reactive flux method and obtained $\Delta V\tst$ for \lith by differentiation.\cite{Rustad2006} This method has also been applied by several other authors.\cite{Kerisit2009,Dang2013,Annapureddy2014} Trajectory analysis of exchange events was used by Sp\aa ngberg \emph{et al.} to deduce an associative exchange mechanism around \lith, which agrees with the results of other studies, without the need for explicit calculation of $\Delta V\tst$.\cite{Spangberg2003a}

Nuclear quantum effects (NQEs), such as zero-point energy and tunnelling, have been shown to affect the structural and dynamical properties of aqueous systems,\cite{Wallqvist1985,Miller2005,Paesani2009,Liu2013} and a natural question is to what extent this is also true for solvation dynamics. Videla \emph{et al.} studied how quantum fluctuations affect the solvent relaxation following a sudden jump in the charge of a neutral solute, as in an electron or proton transfer.\cite{Videla2013} They found that in quantum mechanical simulations, the water relaxed more rapidly. The quantum effect on MRTs around several halide ions was recently investigated by Habershon, who found that the classical $\tau$ was up to 40\% larger than the quantum $\tau$, the ratio decreasing with increasing ion size.\cite{Habershon2014} However, this paper used the rigid-body SPC/E water model, and it has been established that neglecting flexibility and anharmonicity in the O--H bond can lead to an overestimation of the significance of quantum effects.\cite{Habershon2009,Markland2012} A rigid-body water model allows for quantum effects in the intermolecular motion, but it does not allow for quantum effects in the intramolecular motion, which tend to operate in the opposite direction.

A clear example of these competing effects is provided by the self-diffusion in liquid water.\cite{Habershon2009} On one hand, zero point energy in the intermolecular modes of the liquid assists the breaking of hydrogen bonds and speeds up the diffusion. On the other, zero-point energy in the (anharmonic) O--H stretch increases both the average O--H bond length and the molecular dipole moment, strengthening the hydrogen bonds and slowing down the diffusion. The net increase in the diffusion coefficient on including NQEs is therefore smaller for a flexible water model than a rigid-body water model.\cite{Habershon2009}

The results of the competition between intermolecular and intramolecular NQEs can be subtle: Li \emph{et al.} have found that hydrogen bonds weaker than those in pure water are further weakened by NQEs, while stronger hydrogen bonds are strengthened.\cite{Li2011} In terms of water exchange, this leads to the interesting possibility that, if the water-fluoride hydrogen bond is strong enough, quantum effects will strengthen it, making it more difficult for water to leave the first hydration shell of the anion. This would give a quantum mechanical residence time larger than the classical one.

In this paper, we investigate this possibility, and also explore the quantum effect on the MRT in the first coordination shell of a cation, by looking at the water exchange dynamics around \lith and \fluor using classical molecular dynamics and ring polymer molecular dynamics (RPMD).\cite{Habershon2013} The reason for focussing on these two ions is that they are the lightest in their groups and they also have the strongest interactions with their hydration shells: they are therefore the ions for which NQEs are expected to be largest among the alkalis and halides. Having found the quantum effect on $\tau$, we also look at the mechanism of the exchange in the two cases, and investigate the quantum effect on the mechanism by analyzing trajectories.

The remainder of this paper is set out as follows: Section \ref{sec:methods} describes the models we have used for the ion-water interactions, as well as the various methods we have used for calculating the rate  and mechanism of water exchange; Section \ref{sec:results} describes the results of our simulations and discusses the various quantum effects observed; and Section \ref{sec:conclusions} concludes the paper.

\section{Computational Details}\label{sec:methods}

\subsection{Ion-Water Interaction Potentials}\label{sec:potentials}

We used the flexible q-TIP4P/F potential to describe the interactions between water molecules.\cite{Habershon2009} This model is specifically designed for use in path integral simulations, in which it successfully captures several thermodynamic, structural, and dynamical properties of liquid water, including its density, O--O, O--H and H--H radial distribution functions (RDFs), and self-diffusion coefficient.\cite{Habershon2009} The polarizability of the water molecules is not included explicitly in the q-TIP4P/F model, and we chose to treat the ion-water interactions in the same way. As small ions, \lith and \fluor have small polarizabilities,\cite{Mahan1980,*Hattig1998} so a non-polarizable model which takes polarization effects into account in a mean-field manner should be sufficient, especially since our goal here is simply to assess the importance of NQEs in the hydration of the two ions. These effects are rather small, so the use of a simple potential energy function that permits the collection of good statistics is of paramount importance. A polarizable water model would almost certainly be better for making quantitative comparisons with experiments,\cite{Huafeng2002} but that is not our aim here.

There are many non-polarizable ion-water interaction potentials available in the literature, and several different sets of experimental target data to which they have been parameterized.\cite{Smith1994,Jensen2006,Joung2008,Yu2010,Reif2011,Peng2012} Two popular choices of target data are the standard free energy of hydration, \dgstan, and the position $r^{\ast}$ of the first maximum in the ion-oxygen RDF.\cite{Jensen2006,Joung2008,Yu2010,Reif2011} We have used these two pieces of experimental data to parameterize our models, and shall describe how they were calculated for each ion in Sec.~II.B.

The ion-water potentials consist of long-range Coulomb interactions between the ion and the partial charge sites on the water molecules, as well as short-range (van der Waals attraction and Pauli repulsion) interactions between the ion and the oxygen atoms. For \lith, the latter were modelled by a Lennard-Jones potential,

\begin{equation}\label{eq:lennardj}
V_{\text{LJ}}(r) = 4\varepsilon\ionox\left[ \left( \frac{\sigma\ionox}{r}\right)^{12} - \left( \frac{\sigma\ionox}{r}\right)^{6}\right],
\end{equation}

\noindent with $r$ the ion-oxygen distance. The parameters $\varepsilon\ionox$ and $\sigma\ionox$ were based on those of Peng \emph{et al.},\cite{Peng2012} with standard Lorentz-Berthelot combination rules\cite{Allen1987} used to convert the ion-ion interaction parameters in Ref. \onlinecite{Peng2012} into ion-water parameters for the q-TIP4P/F water model. The lithium-oxygen potential thus obtained was found to agree reasonably well with both pieces of experimental data, without requiring any alteration.

On the other hand, for the \fluor ion, we were unable to parameterize the Lennard-Jones potential in such a way that both pieces of data were matched satisfactorily. A softer potential, with $r^{-9}$ dependence instead of $r^{-12}$, gave much better agreement with the target data,

\begin{equation}\label{eqn:6-9}
V_{\text{6-9}}(r) = \frac{27}{4}\varepsilon\ionox\left[ \left( \frac{\sigma\ionox}{r}\right)^{9} - \left(\frac{\sigma\ionox}{r}\right)^{6}\right],
\end{equation}

\noindent where, as in Eqn. \eqref{eq:lennardj}, $\varepsilon\ionox$ is the well depth. To parameterize this model, we began with the Lennard-Jones parameters of Jensen and Jorgensen.\cite{Jensen2006} Starting from the same $\varepsilon\ionox$ and a value of $\sigma\ionox$ that gave the potential minimum at the same ion-oxygen separation as in their potential, both values were varied to give the best agreement with \dgstan\ and $r^{\ast}$. Table \ref{tab:potential} summarizes the resulting ion-water interaction parameters for both ions.

\begin{table}[t]
\caption{\label{tab:potential} Parameters in the present ion-water interaction potentials.}
\begin{tabular}{c | c c c}
\hline\hline
~~Ion~~ & $~~\varepsilon\ionox ~[\text{kcal/mol}]~~$ & $~~\sigma\ionox~[\mbox{\AA}]~~$ & ~~Potential~~ \\
\hline
~\lith & 0.136 & 2.297 & $V_{\text{LJ}}$ \\
~\fluor & 0.054 & 3.791 & $V_{\text{6--9}}$ \\
\hline\hline
\end{tabular}
\end{table}

\subsection{Validation Calculations}

We first calculated various thermodynamic and structural properties of the ionic solutions of \lith and \fluor at infinite dilution, and compared them with experimental values to validate our ion-oxygen potentials. Here, and in the remainder of this paper, both classical and 32-bead path integral simulations were carried out with 215 water molecules and 1 ion, at the experimental density of liquid water. Where temperature was held constant, it was fixed at 298 K using a Langevin thermostatting scheme.\cite{Ceriotti2010} The efficiency of the path integral simulations was enhanced by using ring-polymer contraction for the long-range part of the Coulomb potential.\cite{Markland2008}

In order to calculate $r^{\ast}$, we ran classical molecular dynamics (MD) and path integral molecular dynamics\cite{Parrinello1984} (PIMD) simulations in the canonical ensemble. For the \lith ion, 25 MD simulations of length 2 ns and 25 PIMD simulations of length 2.6 ns were used. For both quantum and classical \fluor, 20 simulations of length 250 ps were carried out. From these, $g\ionox(r)$, the ion-oxygen RDF, and $g\ionhyd(r)$, the ion-hydrogen RDF, were extracted. The first minimum, $r_{\text{\tiny{min}}}$, of $g\ionox(r)$ was then used to calculate the coordination number,

\begin{equation}
n_{\text{\tiny{coord}}} = 4\pi\rho\int_{0}^{r_{\text{\tiny{min}}}}r^{2}g\ionox(r) dr,
\end{equation}

\noindent with $\rho$ the oxygen number density. While $n_{\text{\tiny{coord}}}$ was not explicitly used as a target for parameterization, it did help us to appraise how well our models reproduced the experimental hydration structures of the two ions, as well as the extent to which quantum effects altered these structures.

A raw free energy of hydration, \deltag, was calculated using a thermodynamic integration scheme in which the hydration of the ion was broken down into two stages: in the first, an uncharged atom was ``grown'' in a box of 215 water molecules, and in the second, this atom was charged.\cite{Yu2010} Throughout, the total ion-water interaction could be written as\cite{Horinek2009}

\begin{equation}\label{eq:thermo_int}
V\left(\lambda_{1},\lambda_{2}\right) = \sum_{i=1}^{N_{\text{O}}} \left[ 1-\left(1-\lambda_{1}\right)^{6}\right] V\ionox(r_{\text{I} i}) + \lambda_{2}\sum_{j=1}^{N_{\text{C}}} \frac{q_{\text{I}} q_{j}}{4\pi\epsilon r_{\text{I} j}},
\end{equation}

\noindent where the first sum is over all oxygen atoms, and the second over the charge sites on all water molecules. Here $q_{\rm I}$ is the charge of the ion, $q_{j}$ the charge of site $j$, and $r_{\text{I} j}$ the distance between the ion and site $j$. $\lambda_{1}$ determines the strength of the short-range ion-oxygen potential and $\lambda_{2}$ the strength of the Coulomb interaction, so that $\lambda_{1} = \lambda_{2} = 0$ corresponds to a simulation box without the ion, and $\lambda_{1} = \lambda_{2} = 1$ to a fully hydrated ion.

During the first stage of the integration, $\lambda_{2} = 0$ and $\lambda_{1}$ increases from 0 to 1, and during the second stage, $\lambda_{1} = 1$ and $\lambda_{2}$ increases from 0 to 1. A 12-point Gaussian quadrature was used to calculate the free energy change for each stage,\cite{Horinek2009} giving 24 pairs of values $(\lambda_{1},\lambda_{2})$; for each of these, the system was evolved for 200 ps in the NPT ensemble. The pressure was held constant using the barostatting scheme of Ceriotti \emph{et al.},\cite{Ceriotti2013b} which extends to path integral simulations the scheme proposed earlier by Bussi \emph{et al.}\cite{Bussi2009} The thermodynamic integration for each ion was repeated 5 times to obtain an average value for and standard error in \deltag.

To obtain \dgstan\, from this raw \deltag, we added two correction terms. The first accounts for the fact that \dgstan\, is the free energy difference between ions in the gas phase at standard pressure $p\standstate=1$ bar and in the solution phase at standard molality $b\standstate=1$ mol kg$^{-1}$. This correction, equal to $RT\ln\left(V\standstate_{\text{m,gas}}/V\standstate_{\text{m,sol}}\right) = 1.90$ kcal/mol, is the free energy change when an ideal gas at standard pressure is compressed to the molar volume of the ionic solution.\cite{Lamoureux2006} The second correction, equal to $N_{A} q_{\text{I}} \phi$, accounts for the potential difference $\phi$ on crossing the liquid-vapour interface.\cite{Lamoureux2006,Yu2010} Repeating the analysis in Ref. \onlinecite{Wilson1988} with our water model, we found a surface potential of $-$437 mV at 298 K. This gives a correction of $\pm 10.1$ kcal/mol to \deltag, with the positive sign corresponding to \fluor and the negative sign to \lith. 

\subsection{Rate Theory Calculations}

We followed the procedure outlined by Rey and Hynes to calculate the rate coefficient $k$ for the first-order process of water exchange.\cite{Rey1996} Within the transition state theory approximation, this rate coefficient is given by

\begin{equation}\label{eqn:ktst}
k^{\text{TST}} = \frac{1}{\left(2\pi\mu\beta\right)^{1/2}}\frac{(r^{\ddag})^2 e^{-\beta W(r^{\ddag})}}{\int_{0}^{r^{\ddag}} r^{2} e^{-\beta W(r)} dr},
\end{equation}

\noindent where $r$ is the reaction coordinate (the distance between the ion and the centre-of-mass of the exchanging water molecule), $r\tst$ the transition state separation between the ion-water pair, $\mu$ the reduced mass of this pair, and $W(r)$ the potential of mean force (PMF) along this coordinate. The PMF can be calculated from $g_{\text{\tiny{I--H$_{2}$O}}}(r)$, the ion-water centre-of-mass radial distribution function, as

\begin{equation}\label{eqn:PMF}
W(r) = -\beta^{-1}\ln g_{\text{\tiny{I--H$_{2}$O}}}(r).
\end{equation}

The static simulations we used to find $g\ionox(r)$ and $g\ionhyd(r)$ were also used to find $g_{\text{\tiny{I--H$_{2}$O}}}(r)$, but with one important difference. The correct PIMD expression for a radial distribution function such as $g\ionhyd(r)$ involves an average over the ring polymer beads, whereas the reaction coordinate in RPMD rate theory\cite{Craig2005a} is more conveniently\cite{Craig2005} chosen to be a function of the centroid of the ring polymer.\cite{Collepardo-Guevara2008,Markland2008b} For the quantum-mechanical PMF we therefore calculated $g_{\text{\tiny{I--H$_{2}$O}}}(\bar{r})$, where $\bar{r}$ is the distance between the centroids of the ion and the water molecule. 

While there were no problems with these calculations for the \fluor anion (either classically or quantum mechanically), ions such as \lith for which $g_{\text{\tiny{I--H$_{2}$O}}}(r)$ is very small in the region of $r\tst$ are well-known to suffer from poor sampling of $W(r)$ around the transition state. A large increase in simulation time would be required to overcome this poor sampling, so we opted instead to use thermodynamic integration to calculate $W(r)$ in the region of the barrier.\cite{Spangberg2003a,Guardia1991} In the classical case, we ran simulations with the distance between the \lith ion and the centre of mass of a water molecule constrained between $\mbox{2.2 \AA}$ and $\mbox{3.8 \AA}$, in increments of $\mbox{0.025 \AA}$, and in the path integral case, the same constraints were applied to the centroid distance $\bar{r}$. Constraints were enforced using a method based on RATTLE, but with the single Lagrange multiplier calculated analytically.\cite{Andersen1983} The mean force on the constraint was recorded for each distance, and integrated to give a smooth curve for the PMF in the region of $r\tst$.

Transition state theory neglects dynamical recrossing events, which are generally quite significant in water exchange reactions when the ion-water distance is used as the reaction coordinate.\cite{Rey1996,Spangberg2003a,Rustad2006,Kerisit2009,Dang2013} These events can be taken into account by computing the transmission coefficient $\kappa(t)$:

\begin{equation}\label{eqn:fullrate}
k = \lim_{t\rightarrow\infty} \kappa(t) k^{\text{TST}}.
\end{equation}

\noindent
Classically, we used the method of reactive flux to calculate the transmission coefficient,\cite{CharlesH.Bennett1977,Chandler1978} and quantum mechanically, ring polymer molecular dynamics rate theory.\cite{Craig2005a,Craig2005} The RPMD prescription for $\kappa(t)$ for an $n$-bead ring polymer, which reduces to the classical prescription when $n=1$, is simply\cite{Collepardo-Guevara2008,Markland2008b}

\begin{equation}\label{eqn:kappa}
\kappa(t) = \frac{\left\langle \dot{\bar{r}}(0) \delta\left(\bar{r}(0) - r\tst\right) \theta\left[\bar{r}(t)-r\tst\right]\right\rangle\cent}{\left\langle \dot{\bar{r}}(0)\delta\left(\bar{r}(0) - r\tst\right)\theta\left[\dot{\bar{r}}(0)\right]\right\rangle\cent},
\end{equation}

\noindent where $\theta\left[x\right]$ is the Heaviside step function and $\left\langle\cdots\right\rangle\cent$ denotes a Boltzmann average over the ring polymer phase space at reciprocal temperature $\beta_n=\beta/n$.

Starting configurations for our calculations of $\kappa(t)$ were generated by running constrained simulations in the NVT ensemble. In the classical case, the reaction coordinate $r$ was fixed at $r\tst$, and in the path integral case $\bar{r}$ was fixed at $r\tst$. 15 constrained runs of 8 ns each were carried out, with configurations stored every 4 ps. For each run, the stored configurations were used to start 2000 microcanonical trajectories, along which the transmission coefficient was computed using Eqn. \eqref{eqn:kappa}. We averaged over all 15 runs to find the standard error in the computed rate constant $k$, and hence in the mean residence time $\tau = k^{-1}$.

\subsection{Direct Simulations}

Finally, in order to understand the mechanism of water exchange around \lith and \fluor more fully, we applied the direct simulation method to both systems: classical calculations were run in the NVE ensemble, and path integral simulations with the recently-developed thermostatted RPMD (TRPMD) method.\cite{Rossi2014} In each case, we ran 20 simulations of 4 ns each, and recorded every time an exchange event occurred. We defined an exchange to happen if a water molecule was in the first hydration shell of the ion for 5 ps in the past (except for transient escapes lasting no more than 2 ps), and then spent 95\% of the following 5 ps out of the shell. This meant that only very brief excursions back into the first shell were allowed. For each water molecule that was identified as having left the first shell, we also found the molecule that had entered to replace it, by looking backwards in time and using the same criteria to identify the molecule that was not in the shell before the exchange event.

For every exchange event, the distances of the departing and arriving water molecules from the ion were recorded as a function of time, and these distances were averaged over events to give the (microcanonical) ensemble average behaviour of the exchange. From this we could infer the associative or dissociative nature of the reaction.

These simulations were also used to calculate the mean residence time in the first coordination shell, from a residence correlation function.\cite{Impey1983} For $N$ water molecules, this correlation function was defined in the quantum mechanical calculation as

\begin{equation}\label{eqn:restime}
C_{\text{\tiny{res}}}(t) = \frac{\left\langle \sum_{i=1}^{N}\bar{S}_{i}(0;t^{\ast}) \bar{S}_{i}(t;t^{\ast})\right\rangle\cent}{\left\langle \sum_{i=1}^{N}\bar{S}_{i}(0;t^{\ast})\right\rangle\cent}.
\end{equation}

\noindent Here $\bar{S}_{i}(t;t^{\ast})=1$ if the centroid of the $i^{\text{th}}$ water molecule remains in the first coordination shell of the ion between time 0 and time $t$, except for excursions out of the shell of duration no greater than $t^{\ast}$. This is chosen to be 2 ps, roughly the time that a water molecule takes to diffuse over one molecular radius.\cite{Impey1983} The classical residence correlation function was defined in exactly the same way, but with $n=1$ (i.e., with the centroid of the ring polymer of each atom in the path integral simulation being replaced by the position of the atom in the classical simulation). In both cases (quantum and classical), the MRT $\tau$ was calculated by fitting the long-time decay of $C_{\text{\tiny{res}}}(t)$ to $e^{-t/\tau}$. The MRT thus computed was then compared with the results of our reactive flux calculations.

\section{Results and Discussion}\label{sec:results}

\subsection{Validation of Potential Models}\label{subsec:model_verify}

Table \ref{tab:exp_verify} compares various properties of our ion-water potentials, obtained from classical and path integral calculations, with experimental values. For the \lith ion, the agreement between the results of simulations and those of experiment is reasonably good. Any better agreement is hampered by two obstacles: firstly, as pointed out by Jensen and Jorgensen, $r^{\ast}$ is positively correlated with both $\sigma\ionox$ and $\varepsilon\ionox$, and $\left|\text{\dgstan}\right|$ is negatively correlated with both, precluding the simultaneous fitting of both pieces of data.\cite{Jensen2006} Secondly, we found that altering $\sigma\ionox$ and $\varepsilon\ionox$ to favour one piece of data caused dramatic changes in the coordination number. We therefore used the Lennard-Jones parameters obtained from Ref.~\onlinecite{Peng2012} without adjustment to describe the solvation of lithium ions.

\begin{table}[t]
\caption{\label{tab:exp_verify} Comparison of thermodynamic and structural properties of our ion-water models, from classical and path-integral simulations, with the results of experiment; the standard errors in the final digits are shown in parentheses. \dgstan\, is the standard free energy of hydration, $r^{\ast}$ is the first maximum in the ion-oxygen radial distribution function, and $n_{\text{\tiny{coord}}}$ is the coordination number of the ion.}
\begin{tabular}{r | c c c}
\hline\hline
~~Property~~ & ~~Classical~~ & ~~PIMD~~ & ~~Experiment~~ \\
\hline
\multicolumn{4}{l}{\lith Ion} \\
\hline
$-\text{\dgstan}~$[kcal/mol]~~ & 121.2(1) & 120.0(1) & 126.5(1)\footnote{\label{fn:deltag}Ref. \onlinecite{Tissandier1998}} \\
$r^{\ast} ~[\mbox{\AA}]$~~  & 1.92 & 1.93 & 2.08\footnote{\label{fn:rstar}Ref. \onlinecite{Marcus1988}} \\
$n_{\text{\tiny{coord}}}~~$ & 4.03 & 4.05 & 4\footnote{\label{fn:licoord}Ref. \onlinecite{Varma2006}} \\
\hline
\multicolumn{4}{l}{\fluor Ion} \\
\hline
$-\text{\dgstan}~$[kcal/mol]~~ & 108.4(1) & 109.2(1) & 102.5(1)\footnoteref{fn:deltag}\\
$r^{\ast} ~[\mbox{\AA}]$~~ & 2.63 & 2.63 & 2.63\footnoteref{fn:rstar} \\
$n_{\text{\tiny{coord}}}~~$ & 5.67 & 5.68 & 4--6\footnote{\label{fn:flcoord}Ref. \onlinecite{Ohtaki1993}} \\
\hline\hline
\end{tabular}
\end{table}

For the \fluor ion, we were unable to obtain a satisfactory fit to \dgstan\ and $r^{\ast}$ for the q-TIP4P/F water model using a standard Lennard-Jones potential for the ion-oxygen interaction. But by varying the hard-wall part of the potential, and arriving finally at the $r^{-9}$ dependence in Eqn. \eqref{eqn:6-9}, we were able to match the experimental value of $r^{\ast}$ perfectly while obtaining reasonable agreement (as good as we were able to obtain for Li$^+$) with the experimental \dgstan. This is our justification for using the somewhat unusual 6-9 potential for the interaction between the fluoride ion and the water oxygens.\footnote{A few other publications have used non-polarizable ion-water potentials other than the Lennard-Jones. See, for example, Refs. \onlinecite{Impey1983,Spangberg2003a}.} 

\begin{figure}[t]
\centering
\resizebox{0.9\columnwidth}{!}{\includegraphics{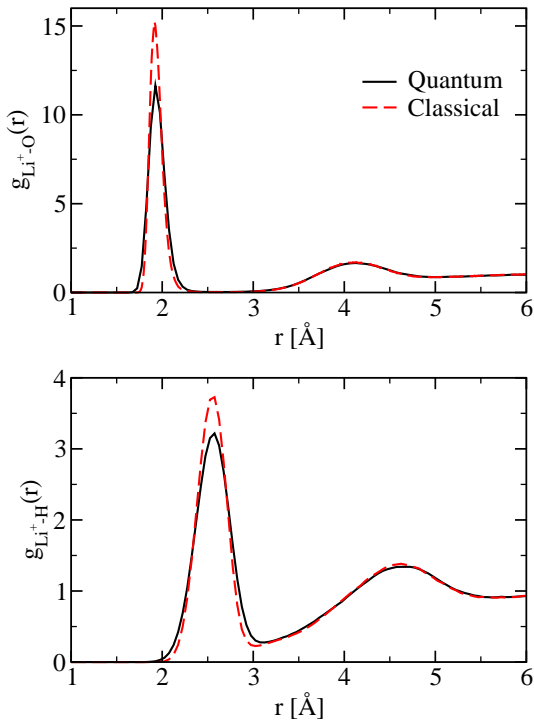}}
\caption{\label{fig:gr_Li} Ion-oxygen (top panel) and ion-hydrogen (bottom panel) radial distribution functions for Li$^+$(aq) from PIMD (solid black lines) and classical MD (dashed red lines) simulations at 298 K.}
\end{figure}

\begin{figure}[t]
\centering
\resizebox{0.9\columnwidth}{!}{\includegraphics{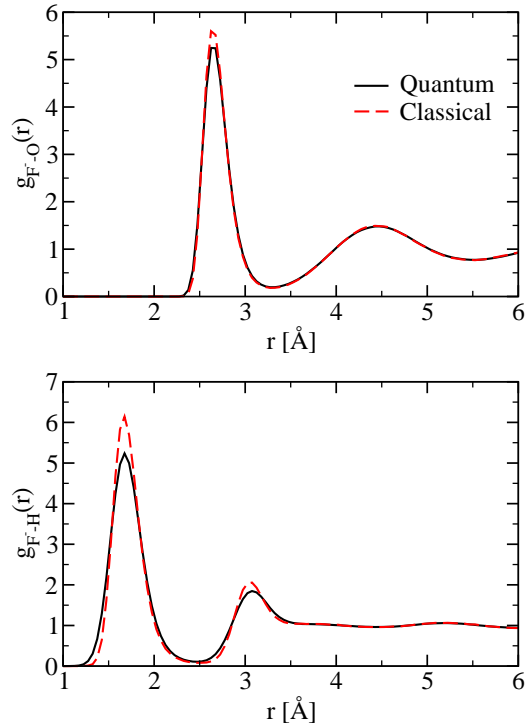}}
\caption{\label{fig:gr_F} Ion-oxygen (top panel) and ion-hydrogen (bottom panel) radial distribution functions for F$^-$(aq) from PIMD (solid black lines) and classical MD (dashed red lines) simulations at 298 K.}
\end{figure}

\begin{figure}[t]
\centering
\resizebox{0.54\columnwidth}{!}{\includegraphics{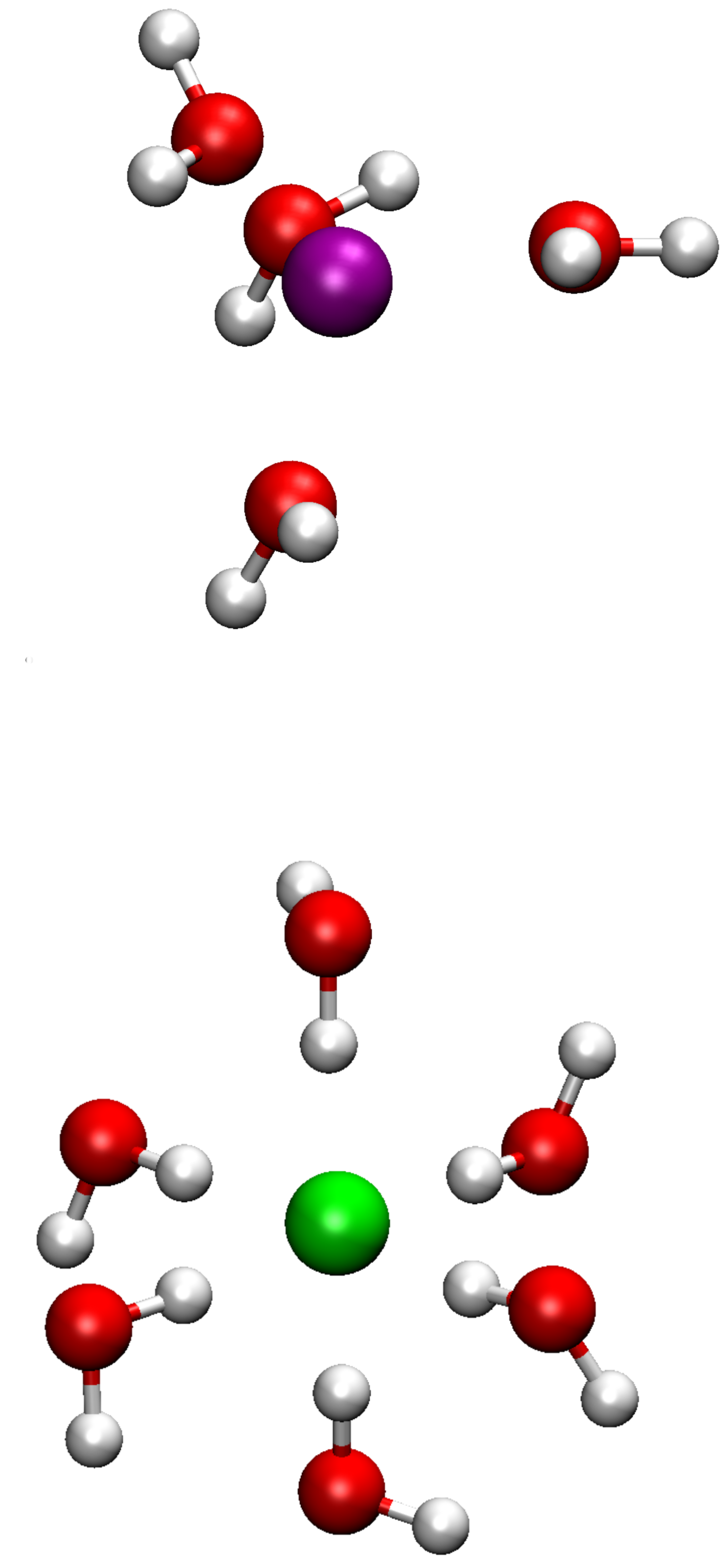}}
\caption{\label{fig:solvation_shell} Snapshots of the first hydration shells of \lith (top) and \fluor (bottom), from classical MD simulations at 298 K.}
\end{figure}

Although the differences between the classical and quantum results for $r^{\ast}$ and $n_{\text{\tiny{coord}}}$ in Table \ref{tab:exp_verify} are negligible, nuclear quantum fluctuations do in fact have a slight effect on the structure of the first hydration shell. Fig.~\ref{fig:gr_Li} shows the ion-oxygen and ion-hydrogen RDFs for \lith obtained from classical MD and PIMD simulations, and Fig.~\ref{fig:gr_F} shows those for \fluor. The structures of the hydration shells are further illustrated by the snapshots in Fig.~\ref{fig:solvation_shell}, taken from our classical MD simulations. The first solvation shell is arranged tetrahedrally around \lith and octahedrally around \fluor, with four water oxygens pointing towards Li$^+$ and six water molecules donating hydrogen bonds to F$^-$. For both ions, the quantum $g\ionox(r)$ is very similar to the classical one, but with a slightly smaller and broader first peak. The same is true for the peaks in $g\ionhyd(r)$ corresponding to the H atoms of first-shell water molecules: for \fluor this is both peaks, and for \lith, the first one. While the quantum effect on $g\ionhyd(r)$ is significantly larger than that on $g\ionox(r)$ for \fluor, the effects on the two RDFs are comparable for \lith.

The decrease in intermolecular structure in quantum simulations of aqueous systems is well-known, and reflects a weakening of the hydrogen-bonding network due to intermolecular zero-point energy.\cite{Kuharski1984,Wallqvist1985} The ion-oxygen RDFs are most relevant to water exchange, and one can see from these that NQEs lead to a slightly less tightly bound first coordination shell. The fact that this effect is more pronounced for \lith than for \fluor is likely due to a combination of two other factors. The first is the mitigation of water-fluoride hydrogen bond weakening by zero-point energy in the O--H bond stretch, which tends to strengthen the hydrogen bond as discussed in Sec.~I. Secondly, the first peak in $g_{{\rm Li}^+{\rm O}}(r)$ is sharper than that in $g_{{\rm F}^-{\rm O}}(r)$. This means that the water molecules are more tightly bound in the first shell of lithium, with more zero-point energy in the coordinates involved in water exchange.\cite{Habershon2014} We will investigate these two factors further in Sec.~\ref{sec:quanteff}.

Turning now to the free energy of hydration, we note that while the difference between our computed quantum and classical values is on the order of 1--2 $k_{\rm B}T$ (which is larger than the uncertainty in these values), it is around 1\% of $\left|\text{\dgstan}\right|$ in each case, representing a very small quantum effect. Overall, quantizing these systems has quite a small effect on their static properties.

\subsection{Rate Theory Results}\label{subsec:ratetheory}

\begin{figure}[t]
\centering
\resizebox{0.9\columnwidth}{!}{\includegraphics{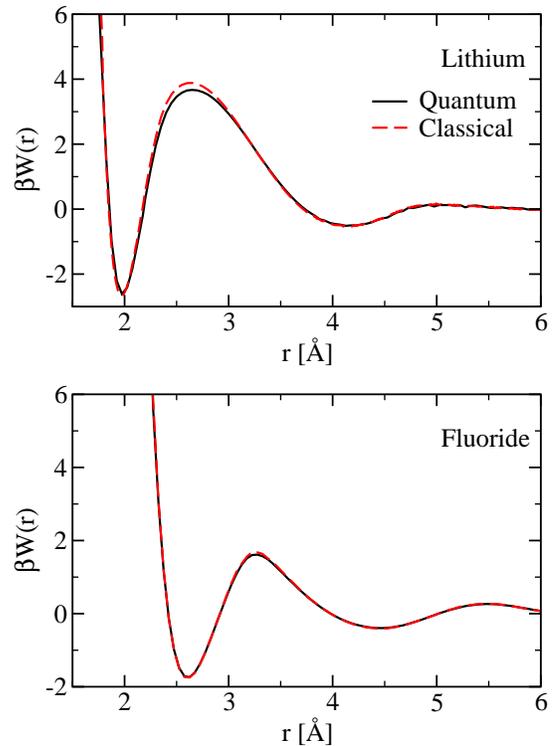}}
\caption{\label{fig:pmf} Potentials of mean force for escape of water from the first solvation shells of \lith and \fluor from PIMD (solid black lines) and classical MD (dashed red lines) simulations at 298 K.}
\end{figure}

In Fig.~\ref{fig:pmf}, we compare the potentials of mean force for escape of water from the first solvation shell, obtained from classical and quantum simulations of both ions. The barrier to exchange of water around \lith in our classical simulation is 6.54 $k_{\text{B}} T$, which is within the range of 5--7 $k_{\text{B}} T$ given in the literature.\cite{Spangberg2003a,Rustad2006,Dang2013} For \fluor, the corresponding barrier is 3.43 $k_{\text{B}} T$. The heights of both barriers are very slightly smaller in the quantum PMFs, by 0.26 $k_{\rm B}T$ for Li$^+$ and by 0.08 $k_{\rm B}T$ for F$^-$. This decrease in the barrier height for F$^-$ is less than half the decrease due to NQEs that Habershon\cite{Habershon2014} found in his calculations with a rigid-body water model (0.2 $k_{\rm B}T$), and we shall attribute this below to the absence of competing quantum effects in these rigid-body simulations.\cite{Habershon2009}

\begin{figure}[t]
\centering
\resizebox{0.9\columnwidth}{!}{\includegraphics{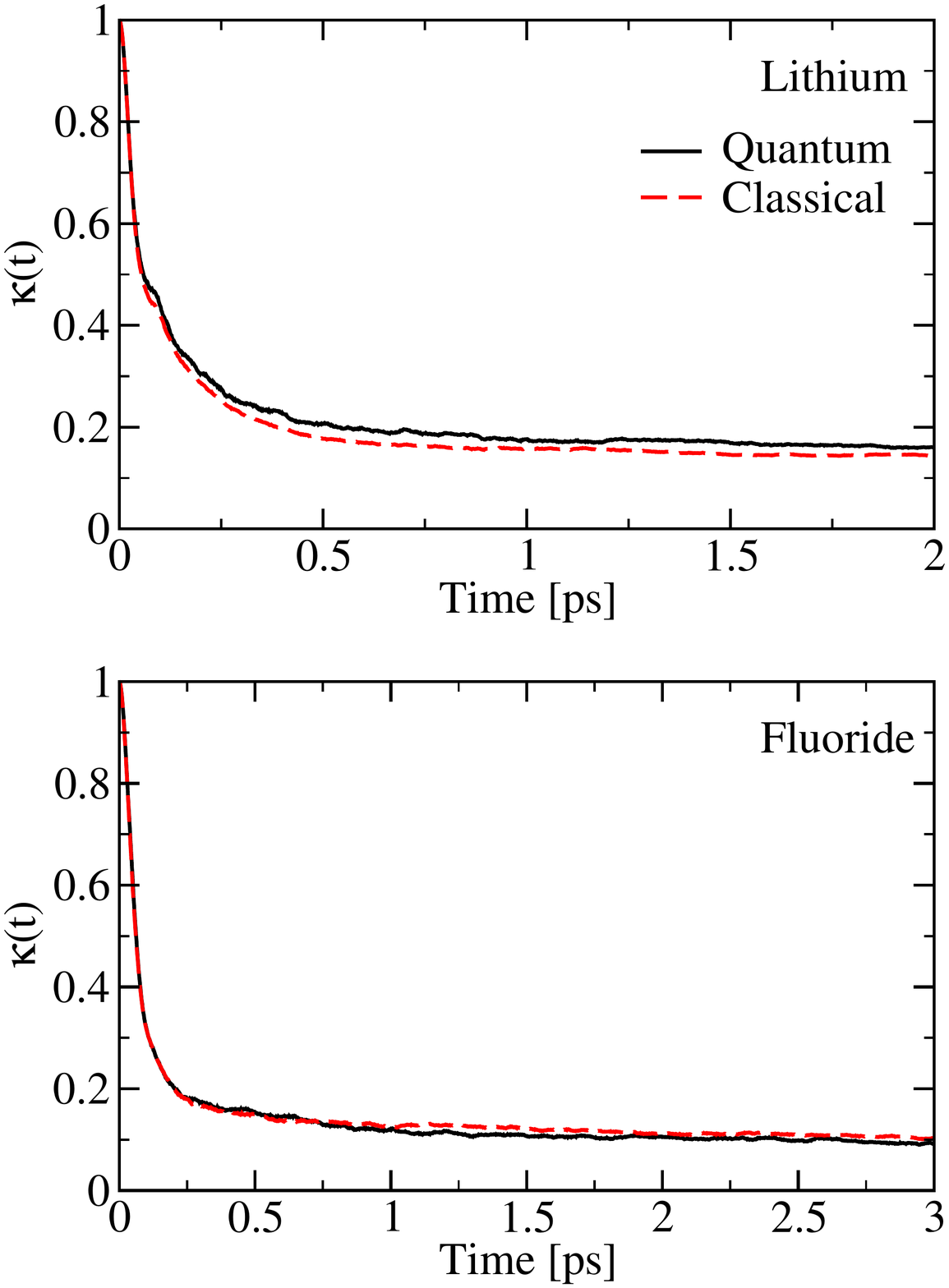}}
\caption{\label{fig:kappas} Transmission coefficients for the water exchange reactions around \lith and \fluor from RPMD (solid lines) and classical MD (dashed red lines) simulations.}
\end{figure}

The transmission coefficients $\kappa(t)$ obtained from our classical MD and RPMD simulations are shown in Fig.~\ref{fig:kappas}. In all cases, the plateau value $\kappa$ is quite small. This implies significant recrossing of the transition state dividing surface, indicating that the ion-water distance $r$ is not the optimum reaction coordinate for these water exchange reactions. It is likely that a better coordinate would be a collective function of the positions of all the water molecules involved in hydration, whose rearrangement is necessary for a molecule to leave the first shell.

Within the computed error bars (which are on the order of the widths of the lines in Fig.~5), the degree of recrossing in the quantum mechanical and classical simulations is the same. Almost all of the nuclear quantum effects on the rate constant are thus captured in the transition-state theory treatment of the reaction, and can be attributed to the weakened hydrogen bonding described in Sec.~\ref{subsec:model_verify}.

\begin{table*}[t]
\caption{\label{tab:mrts} Dynamical properties of water exchange around \lith and \fluor from classical and quantum simulations. $k^{\text{TST}}$ is the transition state theory rate constant, $\kappa$ is the transmission coefficient and $k = k^{\text{TST}}\kappa$ is the full rate constant. $\tau_{\text{\tiny{RF}}} = 1/k$ is the MRT calculated using the reactive flux method and $\tau_{\text{\tiny{DS}}}$ is the decay time of the residence correlation function calculated from direct simulations. $\tau_{\text{cl}}/\tau_{\text{qm}}$ is the ratio of classical to quantum mean residence times. The standard errors in the final digits are shown in parentheses.}
\begin{tabular}{c | c c c c c}
\hline\hline
& $~~k^{\text{TST}}~[$ns$^{-1}]~~$ & $~~\kappa~~$ & $~~k~[$ns$^{-1}]~~$ & $~~\tau_{\text{\tiny{RF}}}~[$ps$]~~$ & $~~\tau_{\text{\tiny{DS}}}~[$ps$]~~$ \\
\hline
\multicolumn{6}{l}{\lith Ion} \\
\hline
Classical~ & 42 & 0.15(1) & 6.2(4) & 160(11) & 160(9) \\
Quantum~ & 51 & 0.16(1) & 8.2(5) & 121(8) & 116(4) \\
& \multicolumn{3}{r}{$\tau_{\text{cl}}/\tau_{\text{qm}}~~$} & 1.34 & 1.38 \\
\hline
\multicolumn{6}{l}{\fluor Ion} \\
\hline
Classical~ & 292 & 0.10(1) & 29.2(2.9) & 34.2(3.4) & 29(1) \\
Quantum~ & 319 & 0.10(1) & 31.9(3.2) & 31.3(3.1) & 26(1) \\
& \multicolumn{3}{r}{$\tau_{\text{cl}}/\tau_{\text{qm}}~~$} & 1.09 & 1.11 \\
\hline\hline
\end{tabular}
\end{table*}

The transition state theory rate constants $k^{\text{TST}}$ for water exchange are given in the first column of Table \ref{tab:mrts}. The quantum rate constant $k_{\text{qm}}^{\text{TST}}$ for each ion is larger than the classical rate constant $k_{\text{cl}}^{\text{TST}}$. For \lith, $k_{\text{qm}}^{\text{TST}}/k_{\text{cl}}^{\text{TST}} = 1.21$, and for \fluor this ratio is 1.09. These results are entirely as predicted in Sec.~\ref{subsec:model_verify}: the water is less strongly bound to the ions in the quantum simulations, the quantum effect being greater for Li$^+$ than for F$^-$.

A similar observation was made by Habershon, who calculated potentials of mean force for exchange around halide ions, and found that a smaller ion-water binding strength was correlated with a smaller decrease in barrier height due to zero-point energy.\cite{Habershon2014} Although the \lith and \fluor ions interact differently with the molecules in their first hydration shells, they appear to fit into this pattern. Since water will be more tightly bound to \lith than to the other alkali metals, and the other alkalis are also heavier than lithium, we would predict that the quantum effects on $k^{\text{TST}}$ seen here are the largest that will be seen in this group.

The quantum and classical transmission coefficients $\kappa$, recrossing-corrected rate coefficients $k = \kappa\,k^{\text{TST}}$, and mean residence times $\tau = k^{-1}$, are also given in Table \ref{tab:mrts}. According to our RPMD calculations, the MRT of water in the first hydration shell of \lith is 118 ps, and in the first shell of \fluor, 31 ps. There are no experimental results with which we can directly compare these values, but neutron scattering experiments suggest that $\tau \lesssim 100$ ps for both ions.\cite{Salmon1987} For \fluor, our results are comfortably within this bound, while for \lith the imprecision in the experimental result makes our value reasonable. The MRTs of \lith from earlier computational studies span two orders of magnitude, with values from 25-400 ps reported,\cite{Szasz1983,Impey1983,Koneshan1998,Spangberg2003a,Rustad2006,Lee1996} while those of \fluor are in the 17-25 ps range.\cite{Heuft2005,Impey1983,Koneshan1998,Habershon2014} 

\subsection{Direct Simulation Results}\label{subsec:directsim}

The mean residence times calculated by fitting the residence correlation function $C_{\text{\tiny{res}}}(t)$ to an exponential decay are given in the final column of Table \ref{tab:mrts}. They compare reasonably well with the results of our rate theory calculations, which gives us confidence in our calculated values. Similar agreement between the two methods was also found by Sp\aa ngberg \emph{et al.}\cite{Spangberg2003a} and by Kerisit and Rosso.\cite{Kerisit2009}

\begin{figure}[t]
\centering
\resizebox{0.9\columnwidth}{!}{\includegraphics{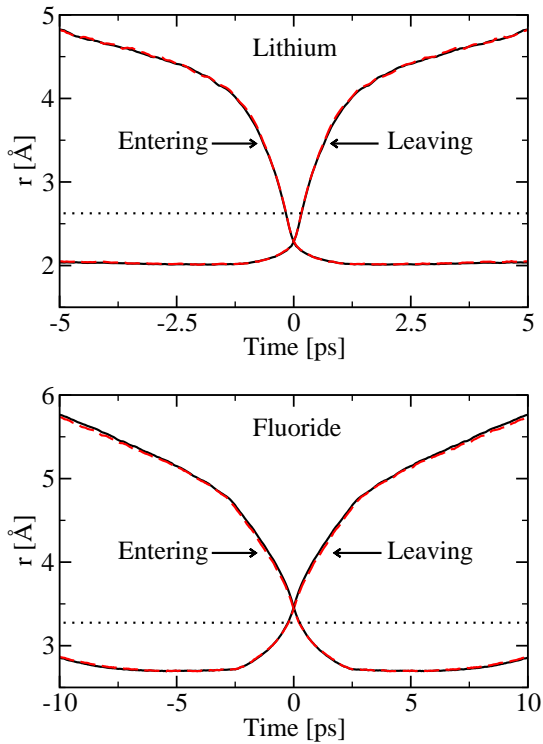}}
\caption{\label{fig:direct_sim} Average trajectories of entering and leaving water molecules during exchange events around \lith and \fluor from TRPMD (solid black lines) and classical MD (dashed red lines) simulations. In each case, the dotted black line shows the position of the first shell boundary.}
\end{figure}

To complement our calculations of $\tau$, Fig.~\ref{fig:direct_sim} shows the average ion-water distance for the leaving water molecule and for its replacement during an exchange event. From this, we can classify the mechanism of exchange around the ion. According to Langford and Gray,\cite{Langford1966} ligand exchanges may be associative, A (in which the initial step is the attachment of the replacing ligand), dissociative, D (in which the initial step is the detachment of a ligand), or interchange, I (with the ligands exchanging in a concerted fashion). This latter mechanism can be further split up depending on whether it has more associative or dissociative character (I$_{\text{a}}$ and I$_{\text{d}}$ respectively).

Using these classifications, we look first at exchange around \lith: the leaving and replacing water molecules cross within the first solvation shell, but the leaving molecule has begun its exit from the shell by the time they cross. This corresponds to an associative interchange (I$_{a}$) mechanism, which agrees with the results of Sp\aa ngberg \emph{et al.},\cite{Spangberg2003a} who found that the majority of exchanges around \lith had an associative nature (either A or I$_{\text{a}}$). The activation volumes calculated in the literature are negative, which also points to an A or I$_{\text{a}}$ mechanism.\cite{Rustad2006,Dang2013}

For \fluor, averaging over all exchange events suggests that the two molecules cross outside the shell, but very close to its boundary, in a dissociative interchange (I$_{\text{d}}$) mechanism. However, Heuft and Meijer found that exchange around \fluor proceeds via two pathways, one of which is associative and the other dissociative.\cite{Heuft2005} To investigate whether the same was true for our model, and the behaviour in Fig.~\ref{fig:direct_sim} was thus an average of two mechanisms, we took each individual event and classified it as associative or dissociative, based on whether the molecules crossed inside or outside of the first shell. 

\begin{figure}[t]
\centering
\resizebox{0.9\columnwidth}{!}{\includegraphics{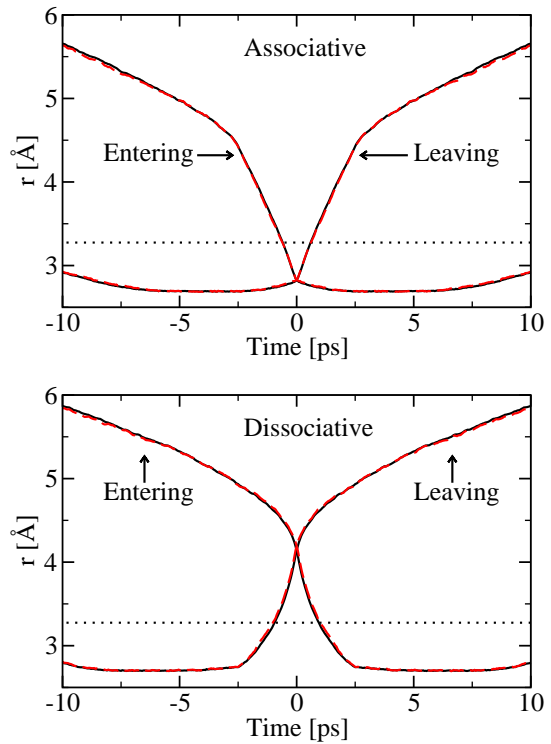}}
\caption{\label{fig:fluoride_dists}  Average trajectories of entering and leaving water molecules during associative and dissociative exchange events around \fluor from TRPMD (solid black lines) and classical MD (dashed red lines) simulations. In each case, the dotted black line shows the position of the first shell boundary.}
\end{figure}

We found that about half of the events were associative, and half dissociative, suggesting that there are indeed two pathways to exchange. In Fig.~\ref{fig:fluoride_dists} we show the distances of the entering and leaving molecules from \fluor, separately averaged over the associative and the dissociative exchanges. The distinction between the two is clear-cut: in both mechanisms, which are classified as I$_{\text{a}}$ and I$_{\text{d}}$ respectively, the molecules cross further from the first shell boundary than in Fig.~\ref{fig:direct_sim}. Repeating the analysis for \lith, we found that fewer than 3\% of the exchanges had any dissociative character, so exchange around this ion can safely be described as associative.

In both Figs.~\ref{fig:direct_sim} and \ref{fig:fluoride_dists}, the results of our TRPMD simulations are compared to those of classical MD. It is evident that, once again, the difference between the quantum and classical simulations is minimal. Coupled with our results from previous sections, we see that the overall effect of quantum fluctuations on water exchange around these two ions is very small. Quantum fluctuations do slightly enhance the rate of water exchange in both cases, by $\sim 10$\% for F$^-$ and $\sim 35$\% for Li$^+$ (see Table~III), but this enhancement can be traced almost entirely to the reduced barrier heights of the quantum mechanical potentials of mean force in Fig.~4. Quantum effects have very little impact on either the time-dependent transmission coefficients in Fig.~5 or the average reactive trajectories in Figs.~6 and~7. Presumably this is because these are determined by the dynamics at configurations on either side of the exchange barrier, where there is hardly any difference between the quantum and classical potentials of mean force (see Fig.~4).

\subsection{Competing Quantum Effects}\label{sec:quanteff}

There is still one remaining aspect of our results that needs explaining. We have shown that nuclear quantum effects decrease the MRT of water in the first coordination shell of \fluor by only $\sim 10$\%, which is considerably smaller than the 40\% decrease found by Habershon.\cite{Habershon2014} We shall now establish that this is because of the competing quantum effects that are present in our simulations: the O--H$\cdots$\fluor hydrogen bond is strengthened by zero-point energy in the anharmonic O--H stretch, and weakened by zero-point energy in perpendicular directions. The first of these effects is missing from the rigid-body water simulations of Ref. \onlinecite{Habershon2014}.

\begin{table*}[t]
\caption{\label{tab:KEtensor} Components of the quantum kinetic energy tensor for H atoms in various environments relative to the \lith and \fluor ions in PIMD simulations at 298 K. The left-hand panel illustrates the tensor, with the directions of its three eigenvectors shown, and $\langle T\rangle$ is the sum of the three eigenvalues.}
\begin{minipage}[c]{0.27\linewidth}
\centering
\centering
\resizebox{\columnwidth}{!}{\includegraphics{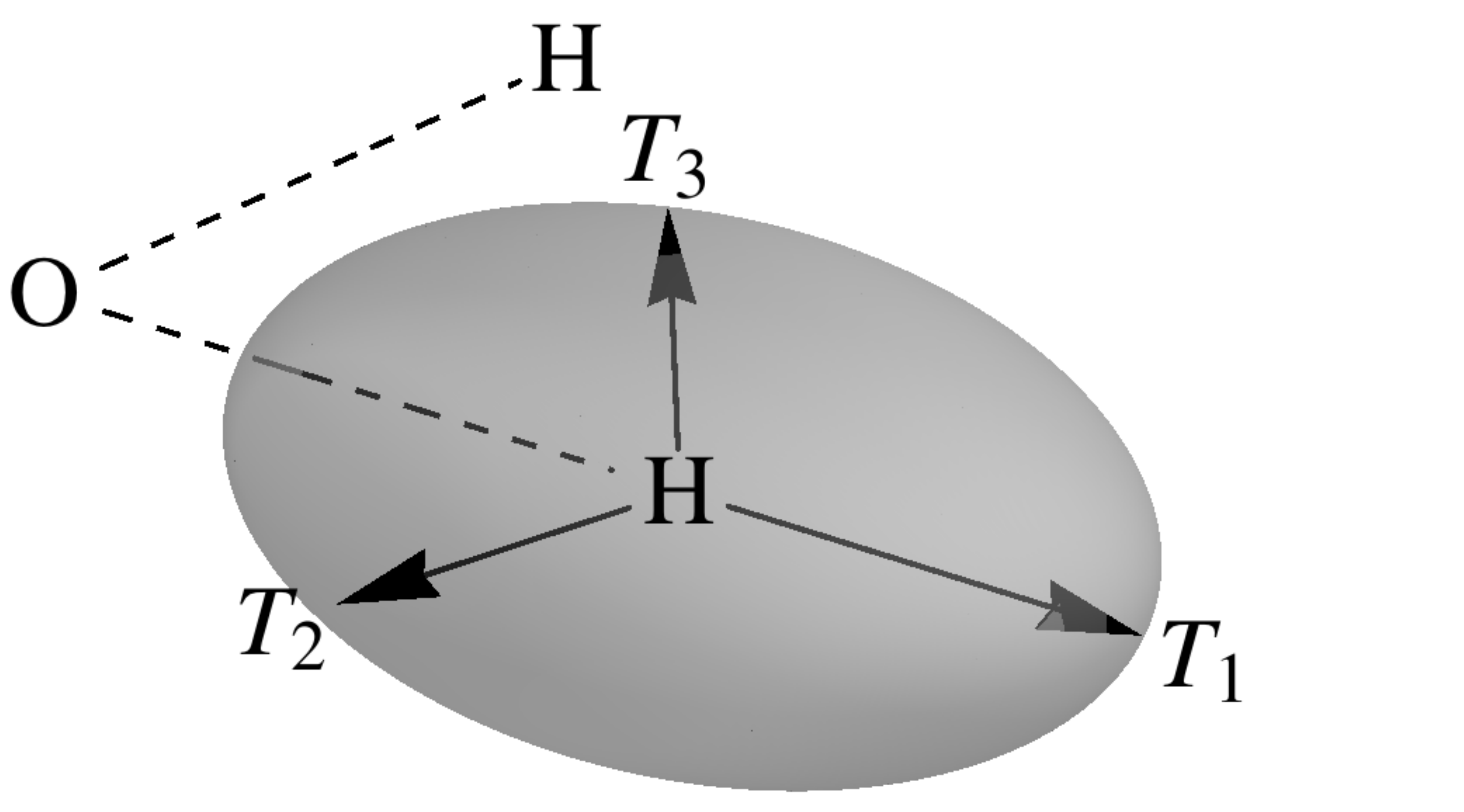}}
\par\vspace{0pt}
\end{minipage}
\begin{minipage}[t]{0.68\linewidth}
\centering
\begin{tabular}{r | c c c c}
\hline\hline
[meV] & $~~\left\langle T_{1}\right\rangle~~$ & $~~\left\langle T_{2}\right\rangle~~$ & $~~\left\langle T_{3}\right\rangle~~$ & $~~~\left\langle T\right\rangle~~~$ \\
\hline
\multicolumn{5}{l}{\lith Ion} \\
\hline
First shell water~ & 98.0 & 34.3 & 21.8 & 154.1\\
Bulk water~ & 98.8 & 33.7 & 21.6 & 154.1\\
\hline
\multicolumn{5}{l}{\fluor Ion} \\
\hline
O--H$\cdots$\fluor hydrogen bonding~ & 91.6 & 38.5 & 28.2 & 158.3 \\
Other first shell hydrogen~ & 99.0 & 33.8 & 21.6 & 154.4 \\
Bulk water~ & 98.9 & 33.7 & 21.5 & 154.1 \\
\hline\hline
\end{tabular}
\par\vspace{0pt}
\end{minipage}
\end{table*}

A convenient probe of competing quantum effects is provided by the quantum kinetic energy $\langle T\rangle$. If this is greater for the H atoms hydrogen-bonded to \fluor than the H atoms in bulk water, the former will be more confined and have a higher zero-point energy than those in the bulk.\cite{Markland2012} What makes this especially useful is that $\langle T\rangle$ can be resolved into contributions from motion in three orthogonal directions, which provides a very natural way to separate the competing quantum effects.\cite{Markland2012}

As in Ref. \onlinecite{Liu2013}, we calculated the centroid virial estimator\cite{Ceriotti2012} for the kinetic energy tensor, $T_{\alpha\beta} = p_{\alpha}p_{\beta}/2 m$, for each H atom in each snapshot of our  PIMD simulation. We then found the rotation that minimized the mean-square difference between the water molecule containing the H atom and a reference molecule, and applied this rotation to the kinetic energy tensor, so as to obtain the tensor in the molecular frame. The resulting tensors were averaged over the simulation for three different groups of H atoms: those in the first hydration shell hydrogen-bonded to \fluor, those in the first hydration shell not hydrogen-bonded to F$^-$, and those in the remainder of the solution (described here as bulk). For each of these species, the averaged tensor $\langle T_{\alpha\beta}\rangle$ was then diagonalised to obtain its eigenvalues $\left<T_i\right>$ and the corresponding eigenvectors.

Table \ref{tab:KEtensor} reports the total kinetic energies $\left<T\right>$ and their components $\left<T_i\right>$ for each of the three types of H atom. In all cases, the components are considerably larger than the classical expectation value of $k_{\rm B}T/2=12.8$ meV, because of the significant amount of zero-point energy in O--H bonds. One sees that the total kinetic energy $\langle T\rangle$ differs appreciably from its bulk value only for those atoms directly hydrogen-bonded to the fluoride ion. $\langle T_{1}\rangle$, the component in the O--H direction, has decreased from its value in bulk water, indicating a decrease in confinement in this direction due to stretching of the bond to facilitate hydrogen bonding. On the other hand, $\langle T_{2}\rangle$ and $\langle T_{3}\rangle$, the components perpendicular to the O--H bond, have increased, reflecting more confinement in these directions due to the stronger hydrogen-bonding interaction of H with the \fluor ion than with the oxygen atoms of other water molecules. This is a clear indication of competing quantum effects. The net result is that  $\langle T\rangle$ is larger by 4.2 meV$\sim k_{\rm B}T/6$ for the H atoms hydrogen bonded to F$^-$ than the H atoms in bulk water. The first solvation shell of the F$^-$ is therefore destabilised slightly by quantum effects, which is consistent with the slight ($\sim 10$\%) reduction in the MRT of the water molecules in this shell we have found in our simulations. Without allowing for the anharmonicity of the O--H bond, the decrease in the $\langle T_{1}\rangle$ component between the bulk and the first solvation shell would be expected to be smaller, resulting in a larger increase in both $\langle T\rangle$ and the exchange rate on including NQEs, as Habershon found in his simulations.\cite{Habershon2014}

Table \ref{tab:KEtensor} also compares the components of $\langle T\rangle$ for H atoms in the first shell of \lith with those in the bulk. The total kinetic energy is the same in both environments, though there is a small difference in the individual components: as for \fluor, $\langle T_{1}\rangle$ is smaller in the first shell, while $\langle T_{2}\rangle$ and $\langle T_{3}\rangle$ are larger. However, in this case the kinetic energy of the H atoms does not play any part in explaining the quantum effect on the exchange rate, since the H atoms do not participate in hydrogen bonding with the ion. Our results for Li$^+$ are in agreement with the results of Videla \emph{et al.},\cite{Videla2013} who found that the spatial delocalization of H atoms in the first solvation shells of cations was very similar to that in bulk water. From the point of view of the H atom kinetic energy, there is certainly very little distinction between the two environments.

The correlation between halide-water binding strengths and quantum effect on MRTs has been attributed to greater zero-point energy in more strongly bound systems.\cite{Habershon2014} Since \lith and \fluor interact differently with their first hydration shells, it is not immediately clear whether or not this factor contributes to the larger quantum effect for the former. For this reason, we have also used transition state theory to look at water exchange around water. This molecule interacts with its first shell as both a hydrogen bond donor and acceptor, so is in this sense intermediate between the two ions. Computing the PMF for this exchange using MD and PIMD simulations with 216 water molecules, we found a barrier height of 1.4 $k_{\text{B}} T$, even smaller than that for \fluor (3.4 $k_{\text{B}} T$). In a similar vein, $k^{\text{TST}}$ is equal to 1.61 ps$^{-1}$ classically and 1.69 ps$^{-1}$ quantum mechanically: this increase of 5\% is also smaller than the 9\% found in Sec.~\ref{subsec:ratetheory} for \fluor. This lends support to the suggestion that, regardless of the mode of solvent-solute binding, more strongly bound solvent molecules lead to a greater nuclear quantum effect on $\tau$.

\section{Conclusions}\label{sec:conclusions}

In this paper, we have studied the impact of nuclear quantum effects on the static and dynamical properties of water in aqueous solutions of \lith and \fluor ions using classical and path integral molecular dynamics techniques. We have found that for both ions, quantization causes a small destabilization of the first hydration shell, leading to a shorter mean residence time for the water molecules in this shell. This is explained in both cases by a net disruption of the hydrogen bonding on the inclusion of zero-point energy. The effect is greater for Li$^+$ than for F$^-$ for two reasons: Li$^+$ interacts more strongly with the water molecules in its first hydration shell, and there is a competition between quantum effects in the O--H$\cdots$\fluor hydrogen bond as discussed in Sec.~III.D. Since an essential ingredient in this competition is the anharmonicity of the O--H bond,\cite{Habershon2009} it is missed in simulations of rigid-body water molecules, which explains the difference between our results for fluoride and those of Ref. \onlinecite{Habershon2014}.

We have also found that the mechanism of water exchange around both ions, exemplified by the trajectories of the leaving and replacing molecules, is practically unchanged by nuclear quantum effects. The real upshot of our results is thus that classical simulations provide a qualitative, and even semiquantitative, description of the water exchange dynamics. Since quantum effects have been seen to diminish with weaker ion-water binding, the \lith and \fluor ions bind water the most strongly in their respective groups, and they are also the lightest elements in their groups, we would expect this to be true for all alkalis and halides. The classical treatment of the nuclear motion in most previous work on water exchange around these ions is therefore justified.

\section*{Acknowledgements}

We would like to thank Anne Wilkins and Josh More for critical reading of the manuscript and helpful comments. D.M.W. acknowledges funding from the Oxford University Clarendon Fund and St.~Edmund Hall, and computer time on the University of Oxford Advanced Research Computing (ARC) facility and the IRIDIS High Performance Computing facility. L.X.D. acknowledges funding from the U.S. Department of Energy, Office of Science, Office of Basic Energy Sciences, Division of Chemical Sciences, Geosciences, and Biosciences.

\end{document}